\pdfoutput=1

\documentclass[11pt]{article}

\usepackage[final]{acl}

\usepackage{times}
\usepackage{latexsym}

\usepackage[T1]{fontenc}

\usepackage[utf8]{inputenc}
\usepackage{enumitem} 
\usepackage{microtype}

\usepackage{inconsolata}
\usepackage{multirow}

\usepackage{graphicx}
\usepackage{subfigure}
\usepackage{booktabs} 

\usepackage{hyperref}

\usepackage{fancyhdr}
\fancypagestyle{firstpage}{
    \fancyhf{}
    \fancyhead[L]{Published as a conference paper at EMNLP 2025}
}
\usepackage{amsmath}
\usepackage{amssymb}
\usepackage{mathtools}
\usepackage{amsthm}

\usepackage[capitalize,noabbrev]{cleveref}

\usepackage{threeparttable}
\usepackage[table]{xcolor}

\theoremstyle{plain}

\theoremstyle{definition}

\theoremstyle{remark}

\newcommand\bsub[1]{\vspace{3pt}\noindent\textbf{#1}}

\usepackage{xspace}
\newcommand{\sysname}{\textsc{EcoLoRA}\xspace}

\title{\sysname: Communication-Efficient Federated Fine-Tuning of Large Language Models}

\author{
 \textbf{Han Liu\textsuperscript{$\spadesuit$}},
 \textbf{Ruoyao Wen\textsuperscript{$\spadesuit$}},
 \textbf{Srijith Nair\textsuperscript{$\clubsuit$}},
 \textbf{Jia Liu\textsuperscript{$\clubsuit$}},
  \textbf{Wenjing Lou\textsuperscript{$\diamondsuit$}},
\\
 \textbf{Chongjie Zhang\textsuperscript{$\spadesuit$}},
 \textbf{William Yeoh\textsuperscript{$\spadesuit$}},
 \textbf{Yevgeniy Vorobeychik\textsuperscript{$\spadesuit$}},
\textbf{Ning Zhang\textsuperscript{$\spadesuit$}},
\\
\textsuperscript{$\spadesuit$}Washington University in St. Louis, \textsuperscript{$\clubsuit$}The Ohio State University, \\ \textsuperscript{$\diamondsuit$}Virginia Polytechnic Institute and State University
 \\
 \small\texttt{\{h.liu1,ruoyao,chongjie,wyeoh,yvorobeychik,zhang.ning\}@wustl.edu}, \\
 \small\texttt{nair.203@osu.edu}, \small\texttt{liu@ece.osu.edu}, 
 \texttt{wjlou@vt.edu}
 }

\begin{document}

\maketitle

\begin{abstract}

To address data locality and privacy restrictions, Federated Learning (FL) has recently been adopted to fine-tune large language models (LLMs), enabling improved performance on various downstream tasks without requiring aggregated data. 
However, the repeated exchange of model updates in FL can result in prohibitively high communication costs, hindering the distributed learning process.  

To address this challenge, we propose \sysname, a novel communication-efficient federated fine-tuning framework for LLMs. Leveraging the modular structure, we propose a round-robin segment sharing scheme, where each client uploads only a complementary LoRA segment per round to reduce network bandwidth. It is further combined with adaptive sparsification methods tailored to LoRA’s training dynamics and lossless encoding techniques. We conduct extensive evaluations on both question-answering and value-alignment tasks across multiple datasets and models. The results show that \sysname significantly reduces communication overhead without compromising performance. For instance, it reduces communication time by up to 79\% and total training time by up to 65\%.

\end{abstract}

\section{Introduction} \label{sec:intro}

With the advancements in scaling laws \cite{kaplan2020scaling}, the parameter sizes of pre-trained language models have grown exponentially \cite{chowdhery2023palm}. Despite this rapid expansion, large language models (LLMs) remain constrained by their inherent knowledge boundaries, limiting their effectiveness in certain downstream tasks \cite{liu2024sequential}. These limitations necessitate task-specific fine-tuning. However, the substantial data required for fine-tuning is often distributed across multiple entities, raising significant privacy concerns related to data sharing.

Federated fine-tuning has emerged as a promising approach to mitigate these concerns. Recent studies have largely focused on integrating parameter-efficient fine-tuning (PEFT) methods into federated learning (FL) to reduce computational costs \cite{che2023federated,cho2024heterogeneous,babakniya2023slora,zhang2024towards,sunimproving,bai2024federated}, where a widely adopted strategy involves transmitting low-rank adaptation (LoRA) modules during the FL process. While LoRA significantly reduces the number of parameters exchanged compared to full fine-tuning, the massive scale of LLMs means that even these modules remain relatively large. Furthermore, repeatedly exchanging these modules during multiple communication rounds results in prohibitively high communication costs, making communication the essential bottleneck in training time.

Such prohibitive overhead can significantly hinder the participation of diverse clients, a key foundation for federated learning. More specifically, network connection speeds and their associated costs vary significantly across different areas, often differing by orders of magnitude \cite{Howdle_2023}. For instance, many less-developed countries achieve bandwidths below 2 Mbps \cite{Husain_2024}, and rural areas often suffer from even poorer connections due to limited infrastructure. 
These disparities can prevent a large percentage of participants from contributing to FL due to expensive and unstable connectivity, excluding valuable high-quality data and undermining fairness in the learning process \cite{dorfman2023docofl}. Furthermore, network speeds are highly asymmetric, with upload speeds often being significantly slower than download speeds \cite{konevcny2016federated}, which presents additional challenges for FL. 

In this work, we propose \sysname, a novel \underline{\textbf{E}}fficient \underline{\textbf{Co}}mmunication framework specifically tailored to the unique training strategies and dynamics of federated fine-tuning of LLMs. First, leveraging the modular structure of LoRA, we introduce a round-robin segment-sharing scheme in which each client transmits only a complementary portion of the LoRA module rather than the entire module. Second, we propose an adaptive sparsification technique customized for the distinct training dynamics observed in matrices A and B of LoRA. Third, the adaptive sparsification method naturally enables parameter distribution suitable for geometric compression, allowing us to employ Golomb coding for further communication efficiency.

To demonstrate the effectiveness of \sysname, we incorporate it into various state-of-the-art methods across different tasks (including general question answering and value alignment), datasets, and models. Our results show significant communication savings while preserving model performance. Notably, \sysname reduces uploaded parameters by up to 89\% and overall communication parameters by up to 58\% compared to existing approaches. Under practical network conditions, it reduces communication time by up to 79\% and total training time by 65\%. Moreover, our approach remains robust under various non-IID data conditions and adds only minimal computational overhead.

Our contributions are summarized as follows:
\begin{itemize}[leftmargin=12pt,itemsep=0pt] 
    \item We propose a novel framework, \sysname, a communication-efficient federated fine-tuning framework for LLMs.
    \item We provide a theoretical proof of the convergence of \sysname.
    \item We conduct extensive experiments, demonstrating that \sysname significantly reduces communication overhead while preserving accuracy.
\end{itemize}

\section{Related Work} \label{sec:related_work}

\subsection{Parameter-efficient Fine-tuning of LLMs}
Fine-tuning is essential for effectively adapting LLMs to diverse domains \cite{zou2024eiven}. However, the sheer scale of LLM parameters renders traditional full-model fine-tuning prohibitively expensive. To address this challenge, various parameter-efficient fine-tuning (PEFT) techniques have been proposed, including prefix-tuning \cite{li2021prefix}, prompt-tuning \cite{lester2021power}, and adapter-based methods \cite{hu2023llm}. Among these approaches, low-rank adaptation (LoRA) \cite{hulora}, which leverages low-rank matrices to re-parameterize pre-trained weight matrices, has received unprecedented attention. LoRA requires tuning less than 1\% of the parameters needed for a full fine-tune while still achieving comparable performance across a wide range of downstream tasks, without introducing additional inference latency. Building on these advantages, numerous LoRA variants have been developed to further improve its efficiency and accuracy \cite{kopiczko2023vera,zhangadaptive,liudora}.

\subsection{Federated Fine-tuning of LLMs} \label{subsec:fed_llm}

Federated learning has attracted substantial research interest \cite{li2025resilient}, serving as a paradigm for addressing data privacy concerns \cite{liu2024please}.
Recently, federated fine-tuning of LLMs has gained significant attention, with most existing studies focusing on integrating PEFT methods into the FL framework to reduce computational costs \cite{che2023federated,wu2024fedbiot,cho2024heterogeneous,babakniya2023slora,zhang2024towards,sunimproving,bai2024federated,liu2024fisher}. For example, \citet{zhang2024towards} incorporates LoRA into the FedAvg framework so that only LoRA modules need to be trained and aggregated. Extending this approach to resource-constrained and heterogeneous scenarios, \citet{wangflora} propose a stacking-based aggregation strategy for heterogeneous LoRA modules, where individual LoRA modules are uploaded for aggregation, and the resulting stacked full-size LoRA weights are distributed back to clients. \citet{sunimproving} further enhances performance under differential privacy guarantees and improves computational efficiency by fine-tuning only the zero-initialized LoRA matrices. Although these approaches reduce both computation and communication costs compared to full fine-tuning, transmitting LoRA modules still imposes considerable overhead. Even though LoRA accounts for a small portion of the total parameters, the massive scale of LLMs means these modules remain large. Repeatedly exchanging them during multiple rounds results in prohibitively high communication costs, making communication the dominant bottleneck in training time.


Another line of research leverages zeroth-order optimization methods for federated LLM fine-tuning \cite{qinfederated,xu2024fwdllm}. While these approaches improve communication efficiency, their reliance on zeroth-order optimization significantly reduces computational efficiency compared to backpropagation-based methods. Consequently, these techniques substantially increase the computation time and prolong the overall training process, particularly in scenarios with limited clients or resource-constrained environments.

\subsection{Communication Optimization in FL}
Communication optimization in traditional federated learning has drawn considerable attention, primarily through three techniques: quantization, sparsification, and client sampling. Quantization methods compress model parameters by representing them with fewer bits \cite{bernstein2018signsgd,leng2018extremely,xu2020ternary,horvoth2022natural}. 
However, quantization typically offers limited compression and may lead to noticeable accuracy degradation, particularly in non-IID settings. Sparsification methods generally achieve higher compression ratios by transmitting sparse representations of model parameters \cite{aji2017sparse,tsuzuku2018variance,sahu2021rethinking}. A representative sparsification technique, top-$k$ sparsification \cite{aji2017sparse}, selects parameters based on magnitude and has demonstrated robustness to non-IID data distributions. Lastly, client sampling approaches selectively include clients based on their expected contributions to model improvement by employing carefully designed criteria  \cite{luping2019cmfl,sun2019communication,tang2022fedcor}.

Federated fine-tuning of LLMs, however, presents new challenges distinct from traditional FL, where PEFT techniques widely adopted in this context could lead to different training dynamics and parameter distributions. As a result, conventional optimization techniques, such as top-$k$ sparsification, may fail to exploit these unique properties, yielding suboptimal communication savings. Similarly, approaches like active client sampling \cite{tang2022fedcor} often incur substantial computational overhead, which undermines their practicality in large-scale LLM fine-tuning scenarios.

\section{Method} \label{sec:method}

\subsection{Problem Formulation}

We consider an FL setting with one server and $K$ devices. Each device $i$ holds a local dataset $\mathcal{D}_i = {(x_j, y_j)}^{n_i}$, where $n_i$, $x_j$, $y_j$  denote the number of samples, the input samples, and labels in client $i$, respectively. The total number of samples across all devices is $N = \sum_{i=1}^{K} n_i$. Following recent state-of-the-art approaches, the pre-trained LLMs $\mathcal{M}$ remain fixed on each device, while only the LoRA parameters are updated and exchanged between device $i$ and the server. 
Suppose $\mathcal{L}(\mathcal{M}, \mathcal{P}, x_j, y_j)$ is the loss evaluated by the model $\mathcal{M}$ with LoRA parameters $\mathcal{P}$ on the local data $(x_j, y_j)$, then the optimization goal is to find a set of LoRA parameters $\mathcal{P}$ to minimize the loss:

\vspace{-2mm}
\begin{equation}
\resizebox{0.95\linewidth}{!}{$
\min_{\mathcal{P}} \quad F(\mathcal{M}, \mathcal{P}, \mathcal{D}) 
= \frac{1}{N} \sum_{i=1}^{K} n_i 
  \,\mathbb{E}_{(x_j,y_j)\sim \mathcal{D}_i} 
\,\bigl[\mathcal{L}(\mathcal{M}, \mathcal{P}, x_j, y_j)\bigr],
$}
\end{equation}

 LoRA models the weight update $\Delta W \in \mathbb{R}^{m \times n}$ through a low-rank decomposition $BA$, where $B \in \mathbb{R}^{m \times r}$ and $A \in \mathbb{R}^{r \times n}$ are two low-rank matrices with $r \ll \min(m, n)$. 
%



\begin{figure}[t]
    \centering
    \includegraphics[scale=0.63]{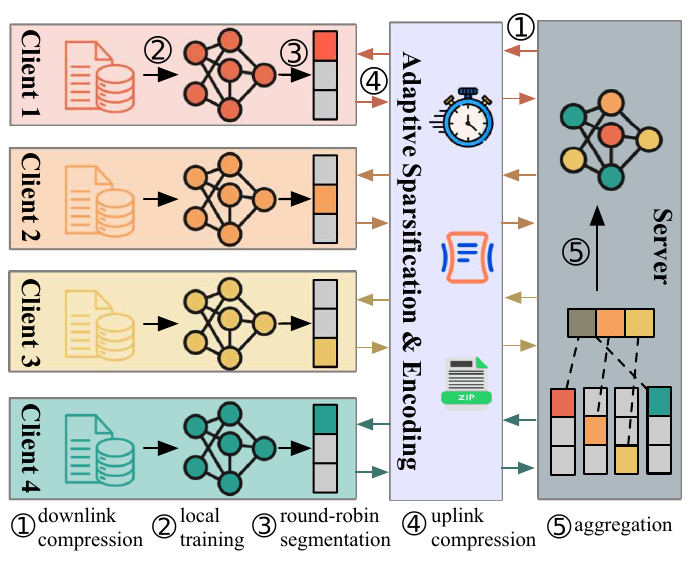}    \caption{Overview of our proposed \sysname.}
    \label{fig:overall_pipeline}
    \vspace{-4mm}
\end{figure}

\subsection{System Model}
Our primary objective is to enhance communication efficiency specifically for federated fine-tuning of LLMs. To guide the design of such methods, we establish the following system goals:

\begin{itemize}[leftmargin=12pt] 
    \item \textit{Communication Efficiency:} The framework should substantially reduce communication overhead while preserving model performance.
    \item \textit{Minimal Computational Overhead:} Since LLM fine-tuning already incurs high computational costs, particularly on resource-constrained edge devices, our framework should introduce minimal additional overhead.
    \item \textit{Robustness to Non-IID Data:} Because data distributions can vary significantly across clients in real-world settings, our framework should remain robust under non-IID conditions. 
\end{itemize}

To address these challenges, we propose \sysname, a novel communication-efficient FL framework illustrated in Figure~\ref{fig:overall_pipeline}. First, we propose a round-robin segment-sharing scheme, leveraging the modular structure of LoRA. Instead of transmitting the entire LoRA module, each client shares only a complementary portion, significantly reducing communication overhead. Second, we introduce an adaptive sparsification tailored for the different training dynamics observed in matrices A and B. This method dynamically compresses parameters based on their training behavior, ensuring minimal performance degradation. Third, the adaptive sparsification method naturally enables a parameter distribution suitable for geometric compression, which we exploit through Golomb coding \cite{golomb1966run} to further optimize communication efficiency.
We elaborate round-robin segment sharing, adaptive sparsification, and encoding in Sections~\ref{subsec:round-robin}, \ref{subsec:sparsification}, and \ref{subsec:encoding}, respectively. Additionally, we analyze computational overhead in Section~\ref{subsec:compution_analysis} and provide convergence analysis in Section~\ref{subsec:convergence_proof}.

\subsection{Round-Robin Segment Sharing} \label{subsec:round-robin}
LoRA can be treated as a modular plug-in to the base model as each LoRA module can be independently attached or removed. Leveraging this modularity, we propose a novel round-robin segment sharing scheme to reduce communication costs, where each client only shares a portion of its LoRA parameters in each round. Formally, we partition the LoRA parameters across all layers into $N_s$ equally sized segments, denoted as $\mathcal{P} = [s_0, s_1, \ldots, s_{N_s-1}]$. In each training round $t$, each client $i$ uploads only one segment, with the ID identified by $(i + t) \bmod N_s$. To ensure that all segments are uploaded by at least one client in each round, enabling complete LoRA parameter updates, we further require $N_s \le N_t$, where $N_t$ is the number of participating clients per round.

At the server side, segments with the same ID are aggregated by a weighted average, and the global LoRA model is reassembled from these aggregated segments. Let $\mathcal{P}^t$ denote the aggregated global LoRA model in the $t$-th round, $s_{i,s}^t$ represent the $s$-th segment uploaded by the $i$-th client in the $t$-th round, $c^k$ denote the set of clients who upload the $k$-th segment, and $n_i$ represent the number of samples in client $i$.  The aggregation rule is:

\vspace{-3mm}
\begin{equation}
\small
\mathcal{P}^t = \left[ \frac{\sum\limits_{i \in c^0} n_i  s_{i,0}^t}{\sum\limits_{i \in c^0} n_i}, \frac{\sum\limits_{i \in c^1} n_i s_{i,1}^t}{\sum\limits_{i \in c^1} n_i}, \dots, \frac{\sum\limits_{i \in c^{N_s-1}} n_i s_{i,N_s-1}^t}{\sum\limits_{i \in c^{N_s-1}} n_i} \right],
\end{equation}

For example, consider $N_t = 5$ clients and $N_s = 3$ segments. In round $t=0$, client $0$ uploads the segment with ID $(0 + 0) \bmod 3 = 0$, i.e., $s_{0,0}^0$; client $1$ uploads $s_{1,1}^0$; client $2$ uploads $s_{2,2}^0$; client $3$ uploads $s_{3,0}^0$; and client $4$ uploads $s_{4,1}^0$. The server then averages $s_{0,0}^0$ and $s_{3,0}^0$ to form the 0-th segment, averages $s_{1,1}^0$ and $s_{4,1}^0$ to form the 1-th segment, and takes $s_{2,2}^0$ for the 2-th segment. Because each client transmits only a single segment in each round, this round-robin segment sharing scheme reduces the upload communication load to $1/N_s$ of the total parameters.

However, this partial update approach introduces a delay for segments that are not uploaded in a given round, which can increase the number of rounds required to converge. To mitigate potential accuracy degradation, we leverage the local model by taking a weighted average of the global and local models at the beginning of each round before optimization. This ensures that even if a segment is not uploaded in a particular round, its previous state still guides local optimization. Moreover, by mixing the globally shared model (the consensus among clients) with the client’s locally fine-tuned model (adapted to its specific data), we improve robustness under non-IID distributions. In cross-device settings, only a subset of clients participates in each round, which may result in some clients remaining idle for many rounds and thus suffering from stale local parameters that potentially hamper global convergence~\cite{xie2019asynchronous} when using the simple average. To address this, we employ an exponential decay weighting~\cite{chen2019communication} to update the local LoRA model:

\vspace{-2mm}
\begin{equation} \label{eq:receive_aggregate}
    \small
    \mathcal{\hat{P}}^{t}_i = (1-e^{-\beta(t-\tau)})\mathcal{P}^{t} + e^{-\beta(t-\tau)} \mathcal{P}_i^{\tau},
\end{equation}

where $t$ denotes the current global round, $\tau$ is the most recent round in which client $i$ participated, and $\beta$ is a hyperparameter balancing staleness.

\subsection{Adaptive Sparsification} \label{subsec:sparsification}

\begin{figure}[t]
  \centering
  \begin{minipage}{0.236\textwidth}
    \includegraphics[width=\linewidth]{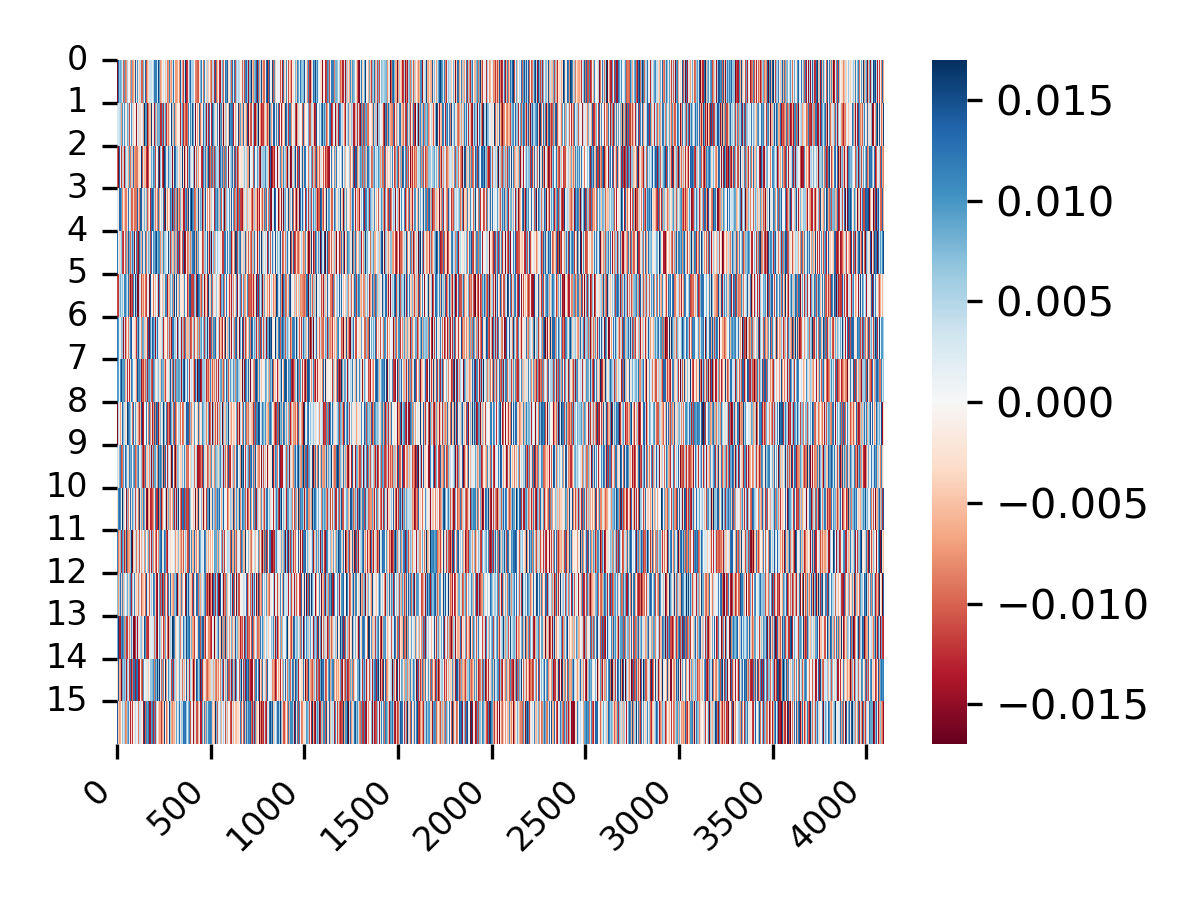}    
    \includegraphics[width=\linewidth]{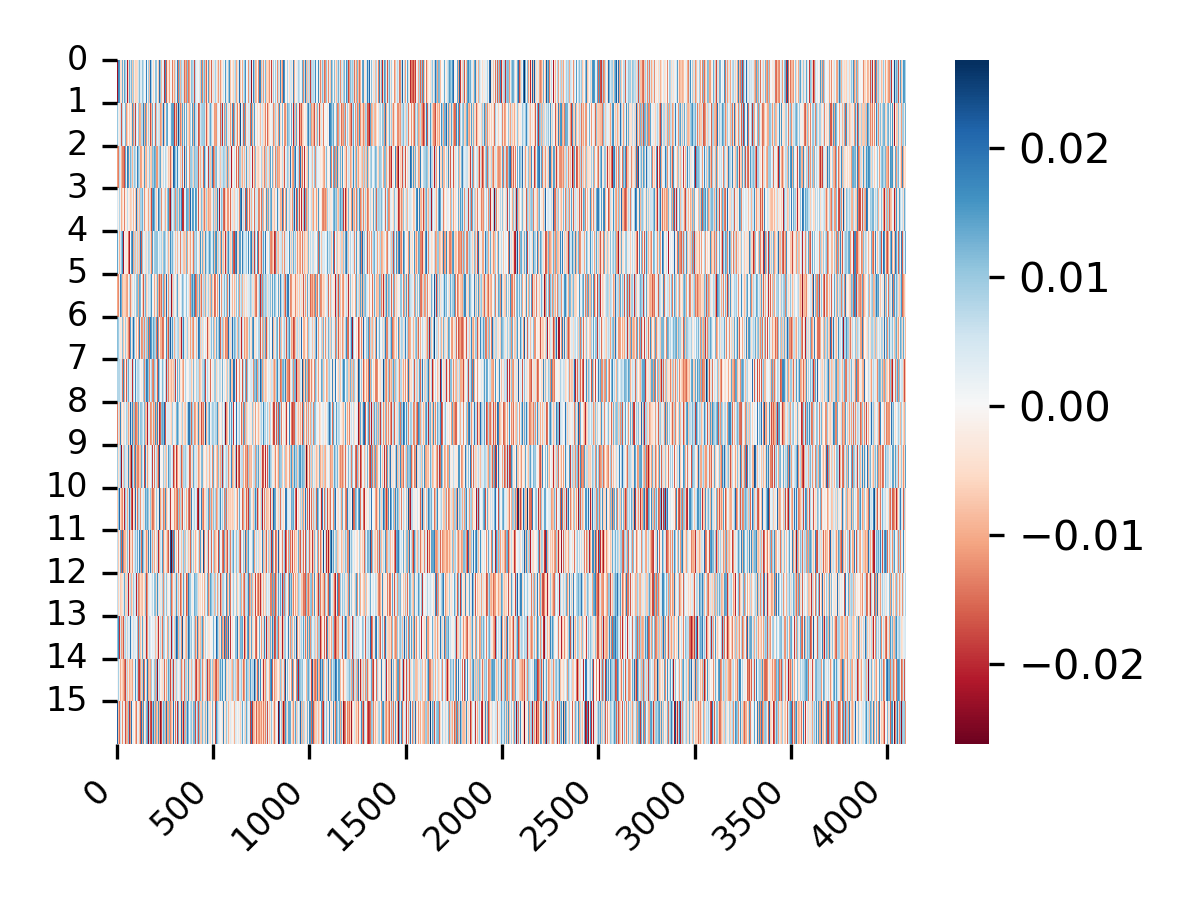}
    \centering
    \text{(a) Matrix A}
  \end{minipage}
  \hfill
  \begin{minipage}{0.236\textwidth}
    \includegraphics[width=\linewidth]{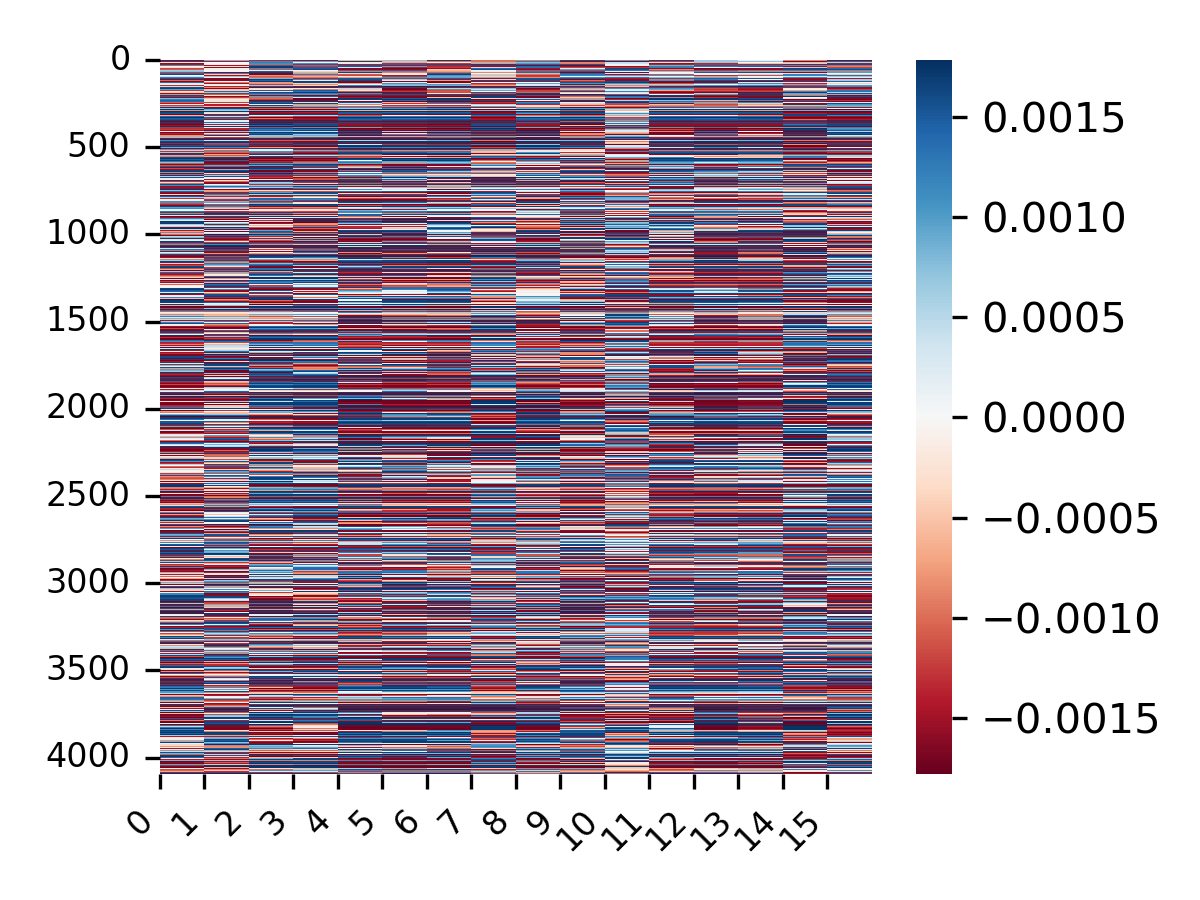}
    \includegraphics[width=\linewidth]{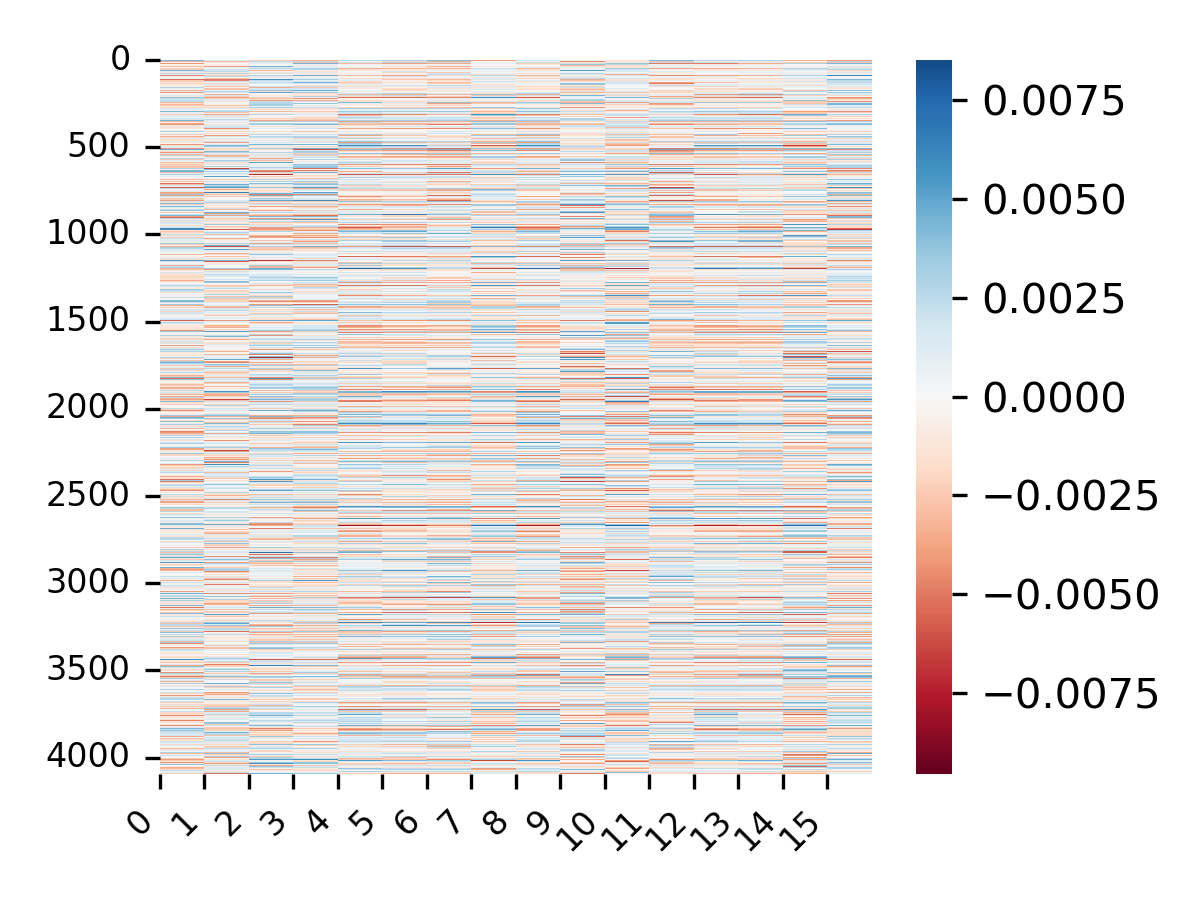}
    \centering
    \text{(b) Matrix B}
  \end{minipage}
  \caption{Visualization of LoRA matrices A and B at epochs 1 (top) and 20 (bottom) during FL training.}
  \label{fig:lora_vis}
  \vspace{-2mm}
\end{figure}

To further reduce the communication load for both uploading and downloading, we can adopt the sparsification techniques that have been successfully applied in traditional FL~\cite{aji2017sparse}. These techniques exploit the observation that most gradient updates are near zero. Among various sparsification approaches, top-$k$ sparsification has demonstrated promising performance with non-i.i.d. data \cite{sattler2019robust} by selecting parameters with the highest $k$ portion of magnitudes for transmission. Since the LoRA module in fine-tuning acts as parameter updates for LLMs, we analyzed matrices A and B during FL training to validate whether the LoRA updates also exhibit similar sparsity in federated LLM fine-tuning, following the experimental setup in Section \ref{subsec:exp_setup}. Figure \ref{fig:lora_vis} shows an example at epoch 0 and epoch 20. Two notable trends emerge from this analysis: (1) As training progresses, both LoRA matrices become sparser, with the many remaining values growing larger in magnitude. (2) Matrices A and B evolve differently; in particular, B becomes much sparser than A. To quantify this, we calculated the Gini coefficient, a statistical measure of distribution inequality where larger values indicate a higher proportion of extreme values. In epoch 0, matrix A had a coefficient of 0.337 and matrix B had 0.243, while by epoch 20, these values reached 0.359 and 0.406 respectively. 
These characteristics present unique opportunities for sparsification. First, to adapt to increasing sparsity, we propose time-adaptive top-$k$ sparsification. We use the loss signal to scale $k$ with training progress, as it both indicates training status and requires no additional computation:

\vspace{-2mm}
\begin{equation}
    \small
    k^t = k_{\min} + (k_{\max} - k_{\min}) \cdot e^{-\gamma( L_0-L_{t-1})},
\end{equation}

where $k^t$ is the sparsity level for round $t$, $L_0$ is the initial loss, $L_{t-1}$ is the global loss for round $t-1$, and $k_{\max}$ and $k_{\min}$ define the sparsification range. As training loss decreases, $k^t$ is reduced, reflecting that the model has learned sufficient knowledge and updates have become sparser. Second, to address the distinct patterns in matrices A and B, we introduce a matrix-adaptive sparsification scheme. We set smaller \(k_{\min}\) value for B (due to its higher sparsity) and use a larger \(\gamma\) for B to capture its rapid change in sparsity.

To mitigate information loss during sparsification, we locally accumulate untransmitted updates as residuals until they become large enough for transmission. Let $\text{SC}_k$ denote top-$k$ sparsification, the compressed parameter $\mathcal{\hat{P}}^{t+1}$ is computed as:

\begin{equation}
    \small
    \mathcal{\hat{P}}^{t+1} = \text{SC}_{k^{t+1}}(\mathcal{P}^{t+1} + R^t),
\end{equation}
where \(R^t\) is the residue at round~\(t\). We then update the residue as:
\begin{equation}
\small
R^{t+1} = R^t + \mathcal{P}^{t+1}- \mathcal{\hat{P}}^{t+1}.
\end{equation}
$R$ is initialized as an empty residual at the beginning of the training. This approach ensures that large updates are transmitted immediately while eventually sending all updates over time.

\subsection{Lossless Encoding} \label{subsec:encoding}
To communicate the set of sparse LoRA tensors between the server and the client, we only need to transmit the positions of the nonzero elements in the flattened tensors, along with a one-bit sign and 16-bit values (assuming FP16) for each nonzero update. However, the positions can still be expensive to communicate because they are typically stored with a fixed number of 16 bits. From an information-theoretic perspective, we can further compress these positions using lossless encoding \cite{sattler2019robust}. 
Specifically, rather than sending the absolute positions of the nonzero elements, we send the distances between consecutive nonzero positions. Given our adaptive sparsification rate $k$, each element is nonzero with probability $k$, thus the distance between two consecutive nonzero elements follows a geometric distribution with parameter $k$, where the probability of a distance of length $n$ is $(1-k)^{n-1}k$. For random variables following a geometric distribution, Golomb coding provides an optimal entropy coding scheme \cite{golomb1966run}. This method represents nonnegative integers using a combination of quotient and remainder, yielding highly compact representations when values follow a geometric distribution. 
For example, when $k=0.1$, using Golomb coding can reduce the average number of bits required to encode each nonzero position to $b^* = 4.8$, which leads to approximately a 3.3$\times$ compression factor per position.

\subsection{Analysis of Computational Overhead} \label{subsec:compution_analysis}
We now analyze the additional computational overhead introduced by our proposed method. For round-robin segment sharing, we compute a weighted average of the global and local models in Eq.~\ref{eq:receive_aggregate}. Since this is an element-wise operation, it requires roughly $2|\mathcal{P}|$ operations. For adaptive sparsification, we could select the top-$k$ LoRA updates using efficient selection algorithms, such as Quicksort, which take about $O\bigl(|\mathcal{P}|\log(|\mathcal{P}|)\bigr)$. Additionally, untransmitted gradients are accumulated as residuals via simple element-wise additions, contributing $O\bigl(|\mathcal{P}|\bigr)$ cost.
For lossless encoding, we first compute the differences between consecutive indices, which takes $O\bigl(k|\mathcal{P}|\bigr)$ time. We then apply Golomb coding to each gap, also running in linear time with respect to $k|\mathcal{P}|$. Overall, the per-round overhead scales nearly linearly with the number of LoRA parameters $|\mathcal{P}|$. Since $|\mathcal{P}|$ is typically much smaller than the full model size $|\mathcal{M}|$, the additional overhead remains minimal compared to the cost of forward and backward propagation.

\subsection{Convergence Analysis} \label{subsec:convergence_proof}

We now present the convergence analysis for \sysname, adhering to the standard procedures described in~\citet{li2019convergence}. Our analysis relies on the following assumptions:

\paragraph{Assumption 1 (Smoothness).} The objective function \(F\) is \(L\)-smooth, meaning:
{\small
\begin{equation*}
F(P_{t+1}) \leq F(P_t) + \langle \nabla F(P_t), P_{t+1} - P_t \rangle + \frac{L}{2}\|P_{t+1} - P_t\|^2.
\end{equation*}
}

\paragraph{Assumption 2 (Bounded Gradients).} The expected squared norm of the stochastic gradients is uniformly bounded by a constant \(G^2\):
{\small
\begin{equation*}
\mathbb{E}\bigl\|\nabla F(P_t)\bigr\|^{2} \leq G^{2}.
\end{equation*}
}

\paragraph{Assumption 3 (Contractive Property).} There exists a constant \(\delta \in (0, 1]\) such that, for any \(x\):
{\small
\begin{equation*}
\|C(x) - x\|^2 \leq (1 - \delta)\|x\|^2.
\end{equation*}
}

We define the following constants:
{\small
\begin{align}
\mu &= \eta\left(\frac{5}{2} + \delta(2\eta L - 1) - 3\eta L\right), \notag \\
\Delta &= \frac{e^{-\beta}}{1 - e^{-\beta}} L^{2}\eta^{2}N_{s}^{2}G^{2}.
\end{align}
}

Under these assumptions, selecting the learning rate within the interval $ \frac{1}{L}\;<\; \eta \;<\;  \frac{5 - 2\delta}{(6 - 4\delta)L}$, after \(T\) communication rounds, \sysname satisfies:
{\small
\begin{equation*}
\frac{1}{T}\sum_{t=0}^{T-1}\bigl\|\nabla F(P_t)\bigr\|^{2}
\leq
\frac{F(P_0) - F^\star}{\mu T}
+
\frac{\eta(2\eta L - 1)\Delta}{\mu}.
\end{equation*}
}

Choosing \(\eta = \mathcal{O}\left(\frac{1}{\sqrt{T}}\right)\), we obtain the final convergence rate:
{\small
\begin{equation*}
\frac{1}{T}\sum_{t=0}^{T-1}\bigl\|\nabla F(P_t)\bigr\|^{2} = \mathcal{O}\left(T^{-1/2}\right)
\end{equation*}
}

The detailed proof is given in Appendix \ref{appendix:proof}.

\begin{table*}[t]
\small
\centering
\begin{threeparttable}
\caption{Comparison of accuracy and associated communication parameters (in millions) across different methods.}
\label{tab:qa_comprehensive}
\renewcommand{\arraystretch}{1.2}
\setlength{\tabcolsep}{4.5pt}
\begin{tabular}{c|c|ccc|ccc}
\hline\hline
\multirow{2}{*}{\textbf{Model}} & \multirow{2}{*}{\textbf{Method}} & \multicolumn{3}{c|}{\textbf{Alpaca}} & \multicolumn{3}{c}{\textbf{Dolly}} \\ 
\cline{3-8} 
                                &                                  & ARC   & Upload Param.  & Total Param.  & ARC   & Upload Param. & Total Param. \\ 
\hline\hline
\multirow{6}{*}{Llama2-7B}      
                                & FedIT                            & 66.6  & 2520.1        & 5040.1        & 66.5  & 2772.1       & 5544.2      \\
                                & \cellcolor{gray!15}FedIT w/ \sysname                  & \cellcolor{gray!15}66.6  & \cellcolor{gray!15}\textbf{346.5} & \cellcolor{gray!15}\textbf{2675.7} & \cellcolor{gray!15}66.5  & \cellcolor{gray!15}\textbf{481.1} & \cellcolor{gray!15}\textbf{3765.6} \\ 
\cline{2-8} 
                                & FLoRA                            & 67.0  & 2856.1        & 31416.9       & 66.4  & 2688.1       & 29568.8     \\
                                & \cellcolor{gray!15}FLoRA w/ \sysname                  & \cellcolor{gray!15}67.2  & \cellcolor{gray!15}\textbf{350.9} & \cellcolor{gray!15}\textbf{24165.7} & \cellcolor{gray!15}66.3  & \cellcolor{gray!15}\textbf{321.6} & \cellcolor{gray!15}\textbf{22023.9} \\ 
\cline{2-8} 
                                & FFA-LoRA                         & 67.4  & 1512.0        & 3024.1        & 66.7  & 1260.0       & 2520.1      \\
                                & \cellcolor{gray!15}FFA-LoRA w/ \sysname               & \cellcolor{gray!15}67.4  & \cellcolor{gray!15}\textbf{160.1} & \cellcolor{gray!15}\textbf{1265.2} & \cellcolor{gray!15}66.7  & \cellcolor{gray!15}\textbf{173.9} & \cellcolor{gray!15}\textbf{1346.1} \\ 
\hline\hline
\multirow{6}{*}{Llama2-13B}     
                                & FedIT                            & 70.3  & 3674.1        & 7348.2        & 70.1  & 2361.9       & 4723.8      \\
                                & \cellcolor{gray!15}FedIT w/ \sysname                  & \cellcolor{gray!15}70.4  & \cellcolor{gray!15}\textbf{488.9} & \cellcolor{gray!15}\textbf{3775.4} & \cellcolor{gray!15}70.0  & \cellcolor{gray!15}\textbf{427.4} & \cellcolor{gray!15}\textbf{3254.8} \\ 
\cline{2-8} 
                                & FLoRA                            & 70.3  & 4461.4        & 49075.3       & 69.8  & 4067.7       & 44745.1     \\
                                & \cellcolor{gray!15}FLoRA w/ \sysname                  & \cellcolor{gray!15}70.5  & \cellcolor{gray!15}\textbf{576.3} & \cellcolor{gray!15}\textbf{39816.7} & \cellcolor{gray!15}70.1  & \cellcolor{gray!15}\textbf{555.8} & \cellcolor{gray!15}\textbf{38026.2} \\ 
\cline{2-8} 
                                & FFA-LoRA                         & 70.2  & 2099.5        & 4199.0        & 69.9  & 2558.7       & 5117.5      \\
                                & \cellcolor{gray!15}FFA-LoRA w/ \sysname               & \cellcolor{gray!15}70.2  & \cellcolor{gray!15}\textbf{272.0} & \cellcolor{gray!15}\textbf{2137.5} & \cellcolor{gray!15}69.9  & \cellcolor{gray!15}\textbf{261.5} & \cellcolor{gray!15}\textbf{1943.3} \\ 
\hline\hline
\end{tabular}
\end{threeparttable}
\vspace{-2mm}
\end{table*}

\section{Experiment} \label{sec:experiment}

\subsection{Experimental Setup} \label{subsec:exp_setup}

\bsub{Models and Datasets. }
We consider two tasks: question answering (QA) and value alignment (VA). For QA, we use Llama2 \cite{touvron2023llama} with 7B and 13B parameters. For VA, we use the uncensored version of Vicuna-7B \cite{xu2023wizardlm}. As instruction datasets for QA, we adopt Databricks-dolly-15k \cite{DatabricksBlog2023DollyV2} and Alpaca-GPT4 \cite{peng2023instruction}. For VA, we use the UltraFeedback dataset \cite{cui2024ultrafeedback}.

\bsub{Evaluation Metrics. }
We measure both model accuracy and communication efficiency. For QA performance, we report results on the ARC easy and challenge benchmark \cite{clark2018think}, taking the average of both sets as the ARC score; for the VA task, we evaluate using MT-bench \cite{zheng2023judging} and MMLU \cite{hendrycks2020measuring} following \cite{wangflora,ye2024openfedllm}. We report communication parameters and time under simulated practical network conditions to assess communication efficiency.

\bsub{Baselines. } Our work proposes a general communication efficient framework to enhance existing federated LLM fine-tuning methods. 
To evaluate its effectiveness, we apply our framework to state-of-the-art approaches: FedIT \cite{zhang2024towards}, FLoRA \cite{wangflora}, and FFA-LoRA \cite{sunimproving}, and compare the resulting performance to the original methods.

\bsub{FL Settings and Implementation Details. } Following \citet{zhang2024towards}, we implement our framework in a federated learning environment with 100 clients. In each round, we randomly sample 10 clients and conduct training for 40 global rounds. To simulate realistic scenarios, we adopt a non-IID data distribution across clients. Detailed experimental configurations and hyperparameter settings are provided in Appendix~\ref{sec:appendix_implementation}.

\subsection{Evaluation Results}

\bsub{Results of QA Tasks. }
Table~\ref{tab:qa_comprehensive} shows the model accuracy on the ARC benchmark and the communication overhead for various methods, both with and without \sysname. Our approach achieves performance comparable to the baseline while significantly reducing communication costs. For example, when applying our method to FFA-LoRA \cite{sunimproving} on Llama2-7B trained with Alpaca, we reduce the required upload communication by 89\%. This reduction is particularly advantageous given that upload speeds are often far slower than download speeds \cite{konevcny2016federated}. Moreover, the total communication parameters are reduced by 58\% under the same setting. Furthermore, the \sysname framework has demonstrated generalizability across different methods, thereby expanding its applicability. For example, it can be combined with approaches that leverage heterogeneous client resources \cite{wangflora} or that strengthen performance under differential privacy constraints \cite{sunimproving}, allowing practitioners to benefit from the respective advantages of each approach.

\bsub{Results of VA Tasks. }  
Alignment with human preferences is a crucial step in LLM post-training \cite{lee2023rlaif}. To evaluate \sysname on this task, we implemented federated direct preference optimization (DPO) \cite{rafailov2023direct} following the approach of \cite{ye2024openfedllm}. Specifically, we use UltraFeedback as our local preference dataset; the response with the highest score is treated as the preferred response, and one of the remaining responses is randomly designated as the dispreferred response, following \cite{tunstall2023zephyr}. As shown in Table~\ref{tab:dpo}, \sysname substantially reduces both the upload and total communication parameters while achieving slightly higher performance on MT-bench and MMLU.

\begin{figure*}[t] 
  \centering
  \subfigure[UL/DL: 0.2/1 Mbps]{
    \includegraphics[width=0.22\textwidth]{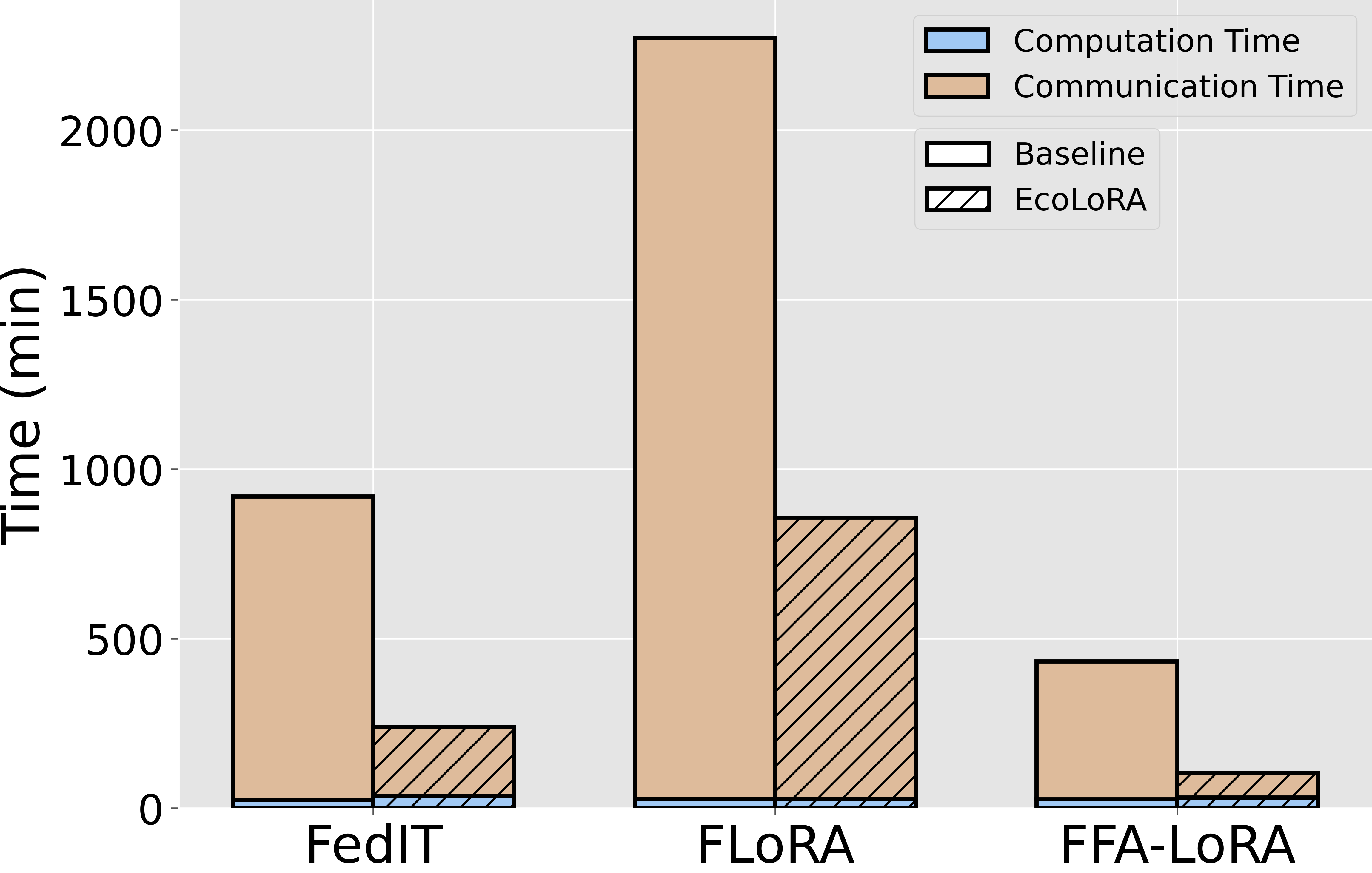}
    \label{fig:up0.2_down1}
  }
  \hfill
  \subfigure[UL/DL: 1/5 Mbps]{
    \includegraphics[width=0.22\textwidth]{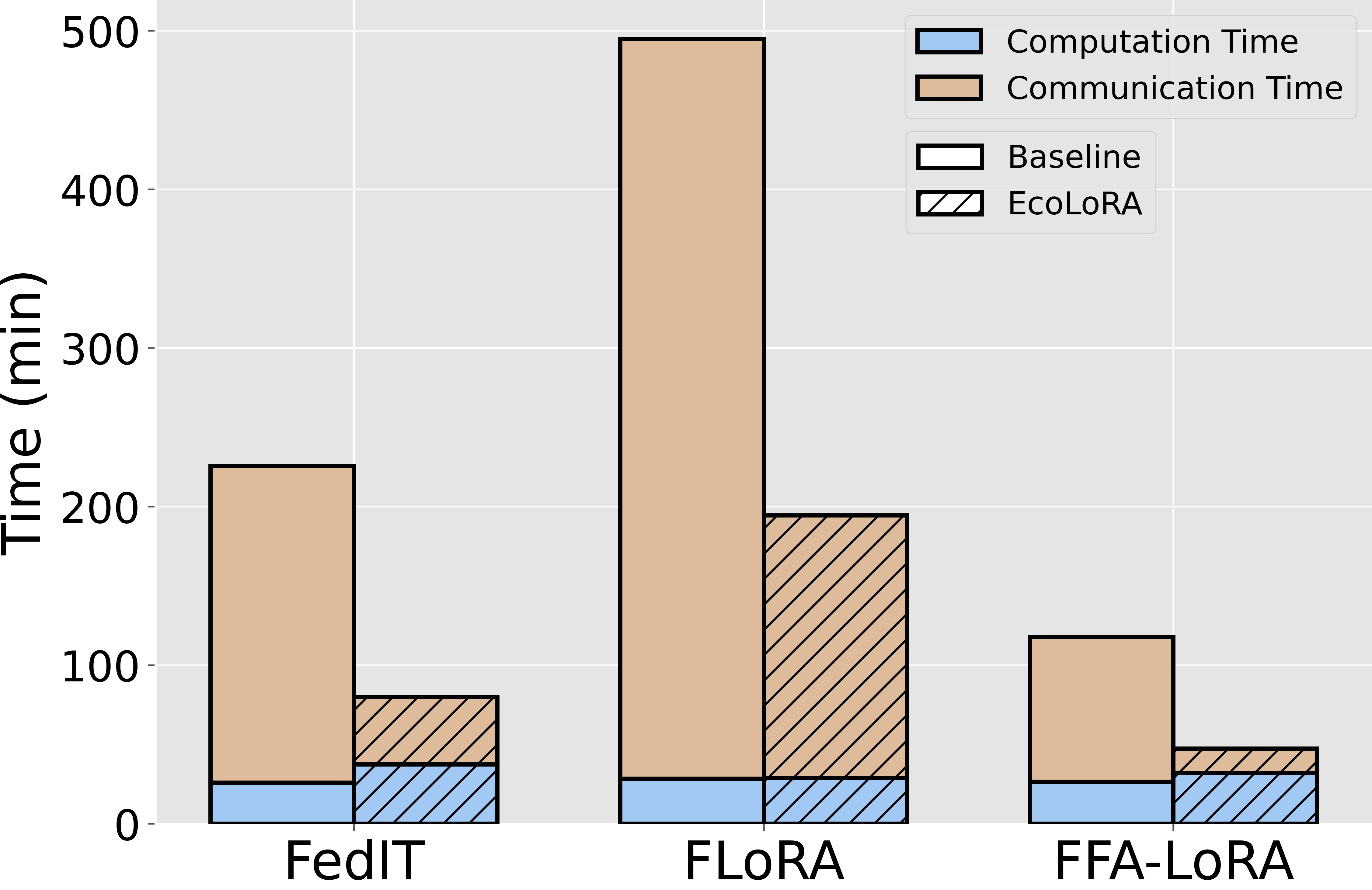}
    \label{fig:up1_down5}
  }
  \hfill
  \subfigure[UL/DL: 2/10 Mbps]{
    \includegraphics[width=0.22\textwidth]{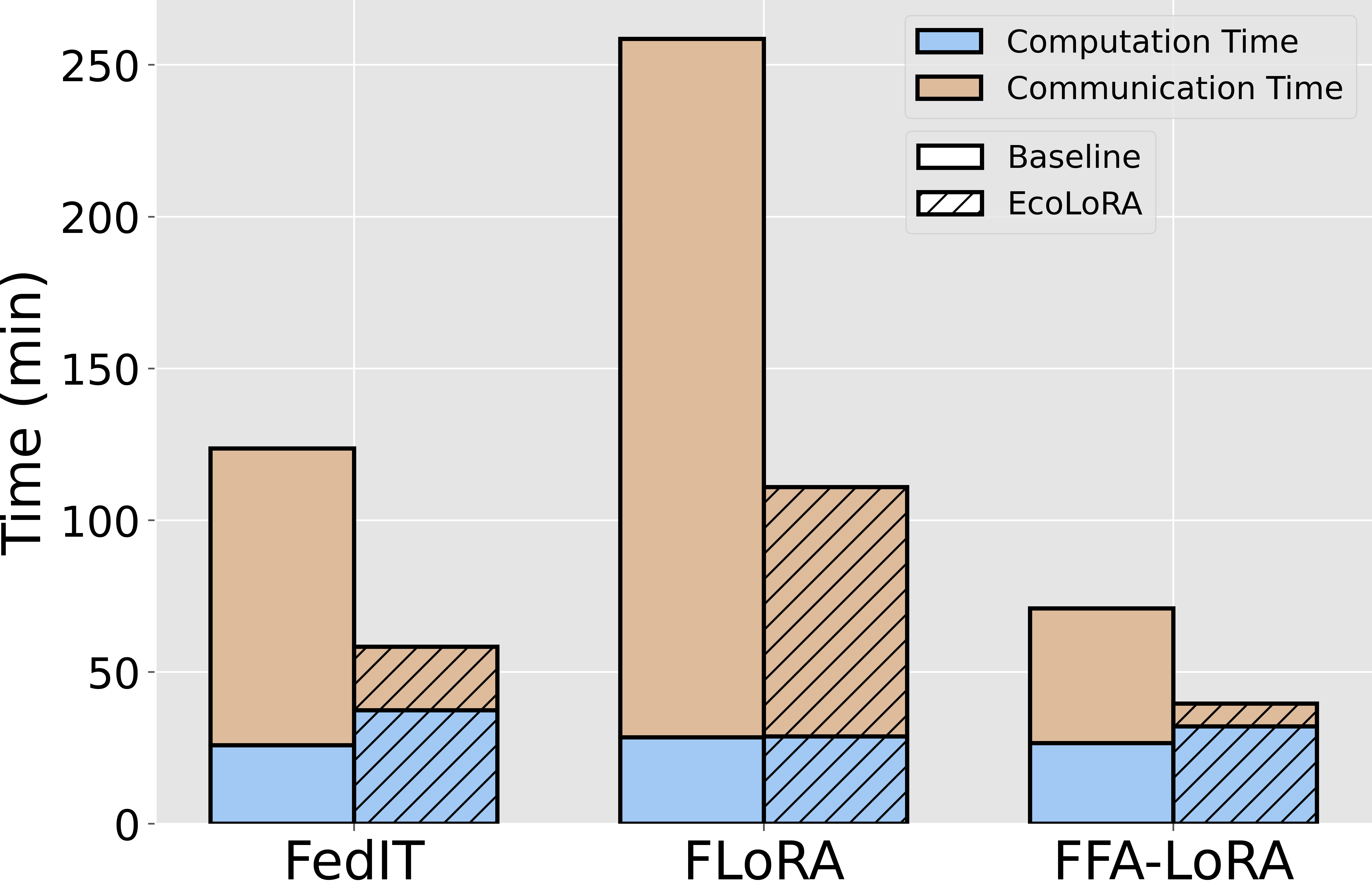}
    \label{fig:up2_down10}
  }
  \hfill
  \subfigure[UL/DL: 5/25 Mbps]{
    \includegraphics[width=0.22\textwidth]{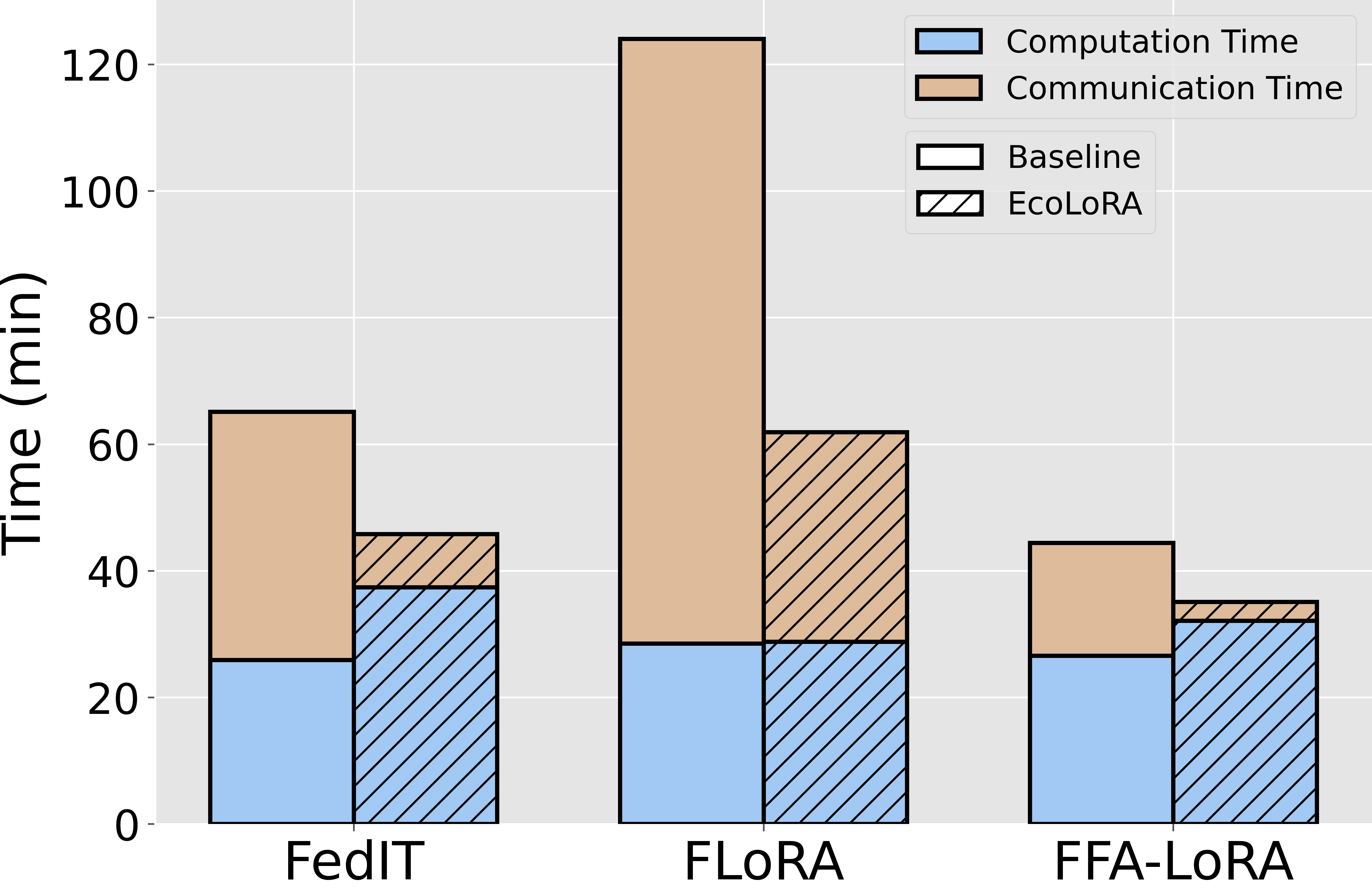}
    \label{fig:up5_down25}
  }
  \caption{The computation and communication time of applying \sysname under different network conditions.}
  \label{fig:time_evaluation}
  \vspace{-3mm}
\end{figure*}

\begin{table}[h]
\small
\centering
\begin{threeparttable}
\caption{Comparison of model accuracy and communication parameters (in millions) of federated DPO with and without \sysname. }
\label{tab:dpo}
\renewcommand{\arraystretch}{1.1}
\setlength{\tabcolsep}{2pt}
\begin{tabular}{c|c|c|c|c}
\hline\hline
\textbf{Method}        & \textbf{MT-bench} & \textbf{MMLU} & \textbf{Upload P.} & \textbf{Total P.} \\ 
\hline\hline
DPO                   & 3.26  & 34.8  & 1719.7   & 3439.3  \\ 
\cellcolor{gray!15}w/ \sysname  & \cellcolor{gray!15}3.28  & \cellcolor{gray!15}35.4  & \cellcolor{gray!15}\textbf{348.8}  & \cellcolor{gray!15}\textbf{2072.1}  \\ 
\hline\hline
\end{tabular}
\end{threeparttable}
\vspace{-3mm}
\end{table}

\subsection{Evaluation in Practical Networks}
To evaluate the performance of \sysname under realistic network conditions, we implemented a simulated federated learning platform following \cite{ekaireb2022ns3}, using \texttt{ns-3}, a widely adopted discrete-event simulator for network communications \cite{henderson2008network}. 
Our simulation adopts \texttt{ns-3}’s point-to-point model to emulate realistic client–server TCP communication. 
We configure the TCP stack to match standard Linux implementations \cite{sarolahti2002congestion}. 

Following practical uplink (UL) and downlink (DL) bandwidth settings in \cite{konevcny2016federated}, we simulate four bandwidth scenarios: 0.2/1 Mbps, 1/5 Mbps, 2/10 Mbps, and 5/25 Mbps, with a fixed latency of 50ms to capture different network conditions. Figure~\ref{fig:time_evaluation} compares the computation and communication time of \sysname against baselines under these scenarios, using Llama2-7B trained on Dolly. Our results demonstrate that as network conditions deteriorate, communication time increasingly dominates the total training time. This effect is particularly notable given that actual throughput typically falls short of theoretical bandwidth; for instance, in our simulated environment, a 1Mbps bandwidth connection achieved an average throughput of only 0.89Mbps. 
These findings underscore the importance of developing communication-efficient fine-tuning methods. 
Across all conditions, \sysname significantly reduces communication overhead while introducing minimal computational cost. For instance, under the 1/5 Mbps setting, it reduces communication time by 79\% and total training time by 65\%. Moreover, the additional per-round computation cost remains \textit{below 3s}, making \sysname a practical solution for resource-constrained environments.

\subsection{Ablation Study} \label{subsec:ablation}

In this section, we analyze the impact of various design components and hyperparameter choices. We present additional experiments in Appendix~\ref{appendix:additional_ablation}.

\bsub{Impacts of Design Components. } 
We conducted an ablation study to investigate how each design component influences both model performance and communication time (both upload and total communication) using Llama2-7B trained on the Dolly dataset with FedIT w/ \sysname method. Specifically, we examine the following variants: (1) w/o Round-Robin (R.R.) Segment: The entire LoRA module is transmitted. (2) w/o Sparsification: The adaptive sparsification method is removed. (3) w/ Fixed Sparsification: A fixed sparsification ratio is used while keeping the overall communication cost identical to that in adaptive sparsification. (4) w/o Encoding: The lossless encoding scheme is excluded. Table~\ref{tab:ablation_components} reports the final accuracy and communication time required to reach the target accuracy of 66.5 for each variant. As shown, each design component notably reduces both the uploading time and total communication time. Additionally using a fixed sparsification ratio results in a significant accuracy drop. This decline occurs because update patterns vary across different training stages and between matrices A and B, which exhibit different levels of robustness to sparsification.

\begin{table}[h]
\small
\centering
\begin{threeparttable}
\caption{Accuracy and communication time for achieving the target accuracy (66.5 on ARC) under different ablations. ("–" indicates target not achieved.)}
\label{tab:ablation_components}
\renewcommand{\arraystretch}{1.1}
\setlength{\tabcolsep}{2pt}
\begin{tabular}{c|c|c|c}
\hline\hline
\textbf{Method}          & \textbf{ARC} & \textbf{Upload Time} & \textbf{Total Time} \\ 
\hline\hline
w/o R.R. Segment         & 66.5         & 72.6                 & 106.2               \\ 
w/o Sparsification       & 66.6         & 25.6                 & 55.6                \\ 
w/ Fixed Sparsification  & 66.1         & -                    & -                   \\ 
w/o Encoding             & 66.5         & 29.5                 & 68.9                \\ 
\cellcolor{gray!15}Full  & \cellcolor{gray!15}66.5  & \cellcolor{gray!15}\textbf{18.2}  & \cellcolor{gray!15}\textbf{42.6}  \\ 
\hline\hline
\end{tabular}
\end{threeparttable}
\end{table}

\bsub{Impacts of Compression Levels. } 
We examine how different compression levels influence model accuracy and communication overhead. In particular, we vary the number of segments \(N_s\) in the Round-Robin scheme, as well as the minimum top-\(k\) thresholds for matrices A and B (\(k_{\min}^A\), \(k_{\min}^B\)), using Llama2-7B trained on the Dolly dataset with FedIT w/ \sysname method. 
Table~\ref{tab:ablation_compress} reports both the accuracy and the communication parameters required to reach a target accuracy under different compression levels. We observe that choosing a smaller \(N_s\) can improve model accuracy and thus reduce download communication overhead (because fewer rounds are needed to achieve the target accuracy). However, it also increases upload communication overhead. Conversely, setting \(N_s\) too large can degrade model accuracy. On the other hand, applying higher sparsity to matrix B than to matrix A (for example, \(k_{\min}^A=0.6\) and \(k_{\min}^B=0.25\)) does not negatively affect model accuracy. As discussed in Section~\ref{subsec:sparsification}, the B matrix is intrinsically sparser than the A matrix. Practitioners should select compression levels achieving an optimal balance between communication costs and accuracy based on the specific network constraints.

\begin{table}[h]
\small
\centering
\caption{Accuracy and communication parameters for achieving the target accuracy (66.5 on ARC) under different compressions. ("–" indicates target not achieved.)}
\label{tab:ablation_compress}
\resizebox{0.48\textwidth}{!}{%
    \begin{threeparttable} 
        \begin{tabular}{c|c|c|c}
        \hline\hline
        \textbf{Method}                          & \textbf{ARC} & \textbf{Upload P.} & \textbf{Total P.} \\ 
        \hline\hline
        \{$N_s=3,k_{\min}^A=0.6,k_{\min}^B=0.5$\}  & 66.6         & 688.9           & 3495.7          \\ 
        \{$N_s=5,k_{\min}^A=0.6,k_{\min}^B=0.5$\}  & 66.5         & 481.1           & 3765.6          \\ 
        \{$N_s=10,k_{\min}^A=0.6,k_{\min}^B=0.5$\} & 66.0         & -               & -               \\ 
        \{$N_s=5,k_{\min}^A=0.6,k_{\min}^B=0.25$\} & 66.5         & 271.2           & 2464.7          \\ 
        \{$N_s=5,k_{\min}^A=0.3,k_{\min}^B=0.5$\}  & 66.2         & -               & -               \\ 
        \hline\hline
        \end{tabular}
    \end{threeparttable}
}
\end{table}

\bsub{Comparison with top-$k$ sparsification. } 
Our proposed adaptive sparsification method exploits the differing sparsity patterns of matrices A and B throughout the training process, in contrast to the fixed threshold used in standard Top-$k$ sparsification. In this section, we present a detailed comparison between the two approaches under varying compression levels. Specifically, we vary the threshold $k$ for Top-$k$ sparsification while ensuring that our adaptive sparsification uses the same total communication budget. The results are shown in Table~\ref{tab:sparsification_comparison}. As shown, while Top-$k$ sparsification achieves comparable performance to our method under low compression, it suffers from performance degradation as the compression level increases. This drop is primarily due to its inability to adapt to the evolving training dynamics and heterogeneous parameter patterns.

\begin{table}[h]
\small
\centering
\begin{threeparttable}
\caption{Comparison of ARC of Top-$k$ and Adaptive Sparsification under varying compression levels.}
\label{tab:sparsification_comparison}
\renewcommand{\arraystretch}{1.1}
\setlength{\tabcolsep}{4pt}
\begin{tabular}{c|c|c}
\hline\hline
\textbf{Threshold $k$} & \textbf{Fixed Top-$k$} & \textbf{Adaptive Sparsification} \\
\hline\hline
0.9 & 66.5 & 66.6 \\
0.7 & 66.1 & 66.5 \\
0.6 & 66.1 & 66.5 \\
0.5 & 65.8 & 66.3 \\
\hline\hline
\end{tabular}
\end{threeparttable}
\end{table}

\bsub{Number of Clients. }
We further examine the impact of scaling the total number of clients. To this end, we evaluate our method with two additional client populations using LLaMA2-7B fine-tuned on Alpaca. The results in Table~\ref{tab:num_clients} show that EcoLoRA consistently reduces communication costs while maintaining accuracy across different client scales.

\begin{table}[h]
\small
\centering
\begin{threeparttable}
\caption{Comparison of accuracy and parameters (in millions) under different numbers of clients.}
\label{tab:num_clients}
\renewcommand{\arraystretch}{1.1}
\setlength{\tabcolsep}{4pt}
\begin{tabular}{c|c|c|c|c}
\hline\hline
\textbf{\# Clients} & \textbf{Method} & \textbf{ARC} & \textbf{Upload P.} & \textbf{Total P.} \\
\hline\hline
200 & FedIT & 66.5 & 3858.8 & 7633.6 \\
200 & w/ EcoLoRA & 66.5 & 501.9 & 2968.2 \\
300 & FedIT & 66.3 & 4529.8 & 8933.9 \\
300 & w/ EcoLoRA & 66.4 & 750.6 & 4450.0 \\
\hline\hline
\end{tabular}
\end{threeparttable}
\end{table}

\bsub{Number of Clients Participating in Each Round. } 
The number of clients participating in each communication round is another critical factor in FL. We fix the total number of clients to 100 and vary the number of participants per round. Experiments with LLaMA2-7B fine-tuned on Dolly (Table~\ref{tab:participate_clients}) show that EcoLoRA remains effective under different participation levels.

\begin{table}[h]
\small
\centering
\begin{threeparttable}
\caption{Comparison of accuracy and parameters (in millions) under varying numbers of participating clients in each round.}
\label{tab:participate_clients}
\renewcommand{\arraystretch}{1.1}
\setlength{\tabcolsep}{3pt}
\begin{tabular}{c|c|c|c|c}
\hline\hline
\textbf{\# P. Clients} & \textbf{Method} & \textbf{ARC} & \textbf{Upload P.} & \textbf{Total P.} \\
\hline\hline
30 & FedIT & 66.5 & 4781.5 & 9437.2 \\
30 & w/ EcoLoRA & 66.6 & 750.5 & 4449.9 \\
50 & FedIT & 66.5 & 4613.7 & 9017.8 \\
50 & w/ EcoLoRA & 66.5 & 645.9 & 3791.6 \\
\hline\hline
\end{tabular}
\end{threeparttable}
\end{table}

\bsub{Number of Local Updates. }
We next study the impact of varying the number of local updates before communication. Experiments with LLaMA2-7B fine-tuned on Dolly (Table~\ref{tab:local_updates}) show that EcoLoRA consistently achieves significant communication reduction while preserving accuracy across different local computation settings.

\begin{table}[h]
\small
\centering
\begin{threeparttable}
\caption{Comparison of accuracy and parameters (in millions) under different numbers of local updates.}
\label{tab:local_updates}
\renewcommand{\arraystretch}{1.1}
\setlength{\tabcolsep}{2pt}
\begin{tabular}{c|c|c|c|c}
\hline\hline
\textbf{\# Local Updates} & \textbf{Method} & \textbf{ARC} & \textbf{Upload P.} & \textbf{Total P.} \\
\hline\hline
3 & FedIT & 66.4 & 1342.2 & 2642.4 \\
3 & w/ EcoLoRA & 66.4 & 151.0  & 880.8 \\
5 & FedIT & 66.3 & 1006.6 & 1971.3 \\
5 & w/ EcoLoRA & 66.4 & 110.7  & 639.2 \\
\hline\hline
\end{tabular}
\end{threeparttable}
\end{table}

\bsub{Impacts of LoRA’s Rank. }
We also investigate the effect of LoRA rank using FFA-LoRA with two different ranks on LLaMA2-7B fine-tuned with Dolly. As shown in Table~\ref{tab:lora_ranks}, EcoLoRA consistently delivers substantial communication savings at both low and high ranks, while maintaining accuracy.

\begin{table}[h]
\small
\centering
\begin{threeparttable}
\caption{Comparison of accuracy and parameters (in millions) under different LoRA ranks.}
\label{tab:lora_ranks}
\renewcommand{\arraystretch}{1.1}
\setlength{\tabcolsep}{3pt}
\begin{tabular}{c|c|c|c|c}
\hline\hline
\textbf{LoRA Rank} & \textbf{Method} & \textbf{ARC} & \textbf{Upload P.} & \textbf{Total P.} \\
\hline\hline
8  & FFA-LoRA & 66.3 & 713.0  & 1405.1 \\
8  & w/ EcoLoRA & 66.3 & 86.6   & 622.5 \\
32 & FFA-LoRA & 66.7 & 2600.4 & 5117.0 \\
32 & w/ EcoLoRA & 66.8 & 333.5  & 2567.5 \\
\hline\hline
\end{tabular}
\end{threeparttable}
\end{table}

\section{Conclusion} \label{sec:conclusion}

In this paper, we introduced \sysname, a novel communication-efficient federated fine-tuning framework for LLMs. Our approach comprises a round-robin segment sharing scheme, an adaptive sparsification method, and lossless encoding. Extensive evaluations on QA and VA tasks across diverse datasets and models show that \sysname substantially reduces communication overhead while maintaining accuracy.
Moreover, it remains robust under non-IID settings and incurs minimal computational overhead.

\section*{Acknowledgment}
We thank the reviewers for their valuable feedback.
This work was partially supported by NSF (CNS-2154930, 
CNS-2229427, 
CNS-2238635, CCF-2403758), 
ARO (W911NF-24-1-0155, W911NF-25-1-0059), 
and ONR (N00014-24-1-2663, N00014-24-1-2730). 

\section*{Limitations}
The primary limitation of this work is that EcoLoRA is developed and evaluated only with LoRA. Although LoRA is currently the most widely adopted PEFT method for federated fine-tuning of LLMs, this focus limits the broad applicability of our method.

On the other hand, we believe the core design principles of EcoLoRA are broadly applicable to other PEFT methods as well. For instance, adapter-based approaches introduce modular trainable layers between frozen backbone parameters, and prefix-tuning leverages independent prefix vectors prepended to transformer blocks. Both of these methods exhibit modularity and layer-wise independence, making them naturally compatible with EcoLoRA’s segment-sharing strategy. We leave a systematic exploration of applying EcoLoRA to these and other PEFT methods to future work.

\section*{Ethical Considerations}
We propose a communication-efficient federated learning framework designed to improve system efficiency while preserving data privacy. Additionally, all our experiments use public datasets, we have not identified any specific risks arising from this study. However, we remain mindful of potential privacy and security implications that may be associated with federated learning in general.


\bibliography{reference}

\appendix
\section{Experimental Settings} \label{sec:appendix_implementation}
The Alpaca-GPT4 dataset contains 52K instruction-following examples generated by GPT-4 using Alpaca prompts. The Dolly dataset consists of 15K text samples created by Databricks employees. The UltraFeedback dataset comprises 64K instructions.

To simulate non-IID data distribution across clients, we divide the datasets using a Dirichlet distribution with $\alpha=0.5$. For the Dolly dataset, we directly use the provided category labels for splitting. Since the Alpaca dataset lacks explicit categories, we generate synthetic ones. Specifically, we concatenate the ‘instruction’ and ‘input’ fields of each sample into a single string, convert these strings into TF-IDF vectors (using up to 1000 features and excluding English stop words), and apply K-means clustering to group samples based on textual similarity. The resulting clusters are treated as synthetic categories, and client-specific datasets are created by applying a Dirichlet-based allocation to these clusters. Additionally, we consider a more heterogeneous non-IID scenario in which each client is assigned data from a distinct task domain.

We set the number of segments $N_s$ to 5 and set the sparsity rates as $k_{\max} = 0.95$, $k_{\min}^A = 0.6$, and $k_{\min}^B = 0.5$. We apply LoRA only to the self-attention layers, following \cite{hulora}. For QA tasks, in accordance with \cite{zhang2024towards,wangflora}, we set the rank $r$ to 16, the scaling factor $\alpha$ to 32, and use a learning rate of $3 \times 10^{-4}$. For VA tasks, following \cite{ye2024openfedllm}, we choose $r=8$, $\alpha=16$, and a learning rate of $5 \times 10^{-4}$. For the Vicuna-7B model, we use an uncensored instruction-following model trained on the filtered WizardLM dataset \cite{xu2023wizardlm}, which does not incorporate human-aligned values. All datasets and models are used strictly for research purposes, in accordance with their respective licenses.
When counting the total communication parameters, we exclude those required to distribute the initial pre-trained LLM. To measure communication time, we repeat each experiment five times and report the average.
Experiments on Llama2-7B are conducted using two NVIDIA GeForce RTX 4090 GPUs, while those on Llama2-13B use an NVIDIA H100 GPU.

\section{Convergence Proof} \label{appendix:proof}

We analyze the convergence of our method following the standard framework adopted in FL literature \cite{li2019convergence}. We assume that the global objective function $F$ is differentiable and $L$-smooth (i.e., its gradient is $L$-Lipschitz continuous).

In each communication round $t$, the global model is updated as:
\begin{equation*}
    P_{t+1} = P_t - \eta U_t,
\end{equation*}
with the effective update given by:
\begin{equation*}
U_t = \nabla F(P_t) + E_t,
\end{equation*}
where $E_t$ contains errors from compression and round-robin segmentation.
By the $L$-smoothness of $F$, we have:
\begin{align*}
\MoveEqLeft F(P_{t+1}) \leq F(P_t) + \langle \nabla F(P_t), P_{t+1} - P_t \rangle \\ &\qquad\quad+ \frac{L}{2} \|P_{t+1} - P_t\|^2.
\end{align*}
By substituting:
\begin{equation*}
P_{t+1} - P_t = -\eta \left(\nabla F(P_t) + E_t\right),
\end{equation*}
we get:
\begin{align} \label{eq:inequality_before}
\MoveEqLeft%
F(P_{t+1}) \leq F(P_t) - \left(\eta - \frac{L \eta^2}{2}\right) \|\nabla F(P_t)\|^2 \nonumber \\
&\underbrace{- \left(\eta - L \eta^2\right) \langle \nabla F(P_t), E_t \rangle}_{\triangleq A} + \frac{L \eta^2}{2} \|E_t\|^2.
\end{align}
Then, using the identity
$$\langle a, b \rangle = \frac{1}{2}\left( \|a\|^2 + \|b\|^2 - \|a - b\|^2 \right),$$
We have:
\begin{align*}
A & = -\eta(1 - \eta L) \langle \nabla F(P_t), E_t \rangle \notag \\
  &= -\frac{\eta}{2}(1 - \eta L)\|\nabla F(P_t)\|^2 \\
  &\quad - \frac{\eta}{2}(1 - \eta L)\|E_t\|^2 \notag \\
  &\quad + \frac{\eta}{2}(1 - \eta L)\|\nabla F(P_t) - E_t\|^2.
\end{align*}

Substituting back into the inequality \eqref{eq:inequality_before}:
\begin{align*}
\MoveEqLeft F(P_{t+1}) 
\leq F(P_t) 
    - \eta\left(\frac{3}{2} - \eta L\right)\|\nabla F(P_t)\|^2 \nonumber \\
&\quad + \eta\left(\eta L - \frac{1}{2}\right)\|E_t\|^2 \nonumber \\
&\quad + \frac{\eta}{2}(1 - \eta L)\|\nabla F(P_t) - E_t\|^2
\end{align*}
Assume $\frac{\eta}{2}(1 - \eta L) < 0 \Rightarrow \eta > \frac{1}{L}$, we have:
\begin{align} \label{inequality_second}
F(P_{t+1}) 
&\leq F(P_t) 
    - \eta\left(\frac{3}{2} - \eta L\right)\|\nabla F(P_t)\|^2 \notag \\
&\quad + \eta\left(\eta L - \frac{1}{2}\right)\|E_t\|^2
\end{align}
Now, we can decompose the error term $E_t$ as:
\begin{equation*}
   E_t = E_t^{\text{comp}} + E_t^{\text{segment}}, 
\end{equation*}
where $E_t^{\text{comp}}$ denotes the adaptive compression error, and $E_t^{\text{segment}}$ denotes the segment sharing error. 
We denote the adaptive sparsification operator as $C(\cdot)$, which satisfies a contractive property, that is, for any vector $x$, there exists a constant $\delta \in (0, 1]$ such that:
\begin{equation*}
\|C(x) - x\|^2 \leq (1 - \delta)\|x\|^2.
\end{equation*}
Then, we get the following bound on the error $E_t^{\text{comp}}$:
\begin{equation*}
\|E_t^{\text{comp}}\|^2 \leq (1 - \delta)\left\|\nabla F(P_t)\right\|^2.
\end{equation*}

In our algorithm, each client updates only one segment per round. Thus, a specific segment only gets updated once every $N_s$ rounds. We denote by $P_t$ the current global parameters and $P_\tau$ the stale parameters from the last round a given client participated. Then by the $L$-smoothness property, we have:
\begin{equation*}
    \|\nabla F(P_t) - \nabla F(P_\tau)\| \leq L\|P_t - P_\tau\|.
\end{equation*}
Since the change in parameters over each round is on the order of the learning rate $\eta$ times the gradient, which we assume is bounded by some $G$, we can get:
\begin{equation*}
\|\nabla F(P_t) - \nabla F(P_\tau)\| \leq L \, \eta \, N_s \, G.
\end{equation*}
As our algorithm uses an exponential decay weighting when updating the local model, we have:
\begin{equation*}
\|E_t^{\text{segment}}\|^2 \leq \sum_{j=1}^{N_s} e^{-\beta j} \cdot \left(L \eta N_s G\right)^2.
\end{equation*}
Because the sum $\sum_{j=1}^{N_s} e^{-\beta j}$ is a geometric series that converges to $\frac{e^{-\beta}}{1 - e^{-\beta}}$, we obtain a bound of the form:
\begin{equation*}
\|E_t^{\text{segment}}\|^2 \leq \frac{e^{-\beta}}{1 - e^{-\beta}} L^2 \eta^2 N_s^2 G^2.
\end{equation*}
We define $\Delta=\frac{e^{-\beta}}{1 - e^{-\beta}} L^2 \eta^2 N_s^2 G^2$, we have:
\begin{align*}
\|E_t\|^2 
&\leq 2 \|E_t^{\text{comp}}\|^2 + 2 \|E_t^{\text{segment}}\|^2 \\
&= 2(1 - \delta)\|\nabla F(P_t)\|^2 + 2 \Delta
\end{align*}
Substituting into the inequality \eqref{inequality_second}:
\begin{align*}
& F(P_{t+1}) \leq F(P_t) + \eta\left(2\eta L - 1\right)\cdot \Delta \\
&  - \eta\left(\frac{5}{2} + \delta(2\eta L-1)-3\eta L \right)\|\nabla F(P_t)\|^2 
\end{align*}
We define $\mu=\eta(\frac{5}{2} + \delta(2\eta L-1)-3\eta L)$, then:
\begin{equation*}
\begin{aligned}
\mu\|\nabla F(P_t)\|^2 
&\leq F(P_{t}) - F(P_{t+1}) \\
&\quad + \eta\left(2\eta L - 1\right)\cdot \Delta
\end{aligned}
\end{equation*}

Summing both sides over $t=0$ to $T-1$:
\begin{equation*}
\begin{aligned}
\sum_{t=0}^{T-1} \mu \|\nabla F(P_t)\|^2 
&\leq F(P_0) - F^* \\
&\quad + T\eta\left(2\eta L - 1\right)\cdot \Delta
\end{aligned}
\end{equation*}

Finally, assuming $\mu > 0 \Rightarrow \eta < \frac{5-2\delta}{(6-4\delta)L}$, we have:
\begin{equation*}
\begin{aligned}
\frac{1}{T} \sum_{t=0}^{T-1} \|\nabla F(P_t)\|^2 
&\leq \frac{F(P_0) - F^*}{\mu T} \\
&\quad + \frac{\eta\left(2\eta L - 1\right) \Delta}{\mu} 
\end{aligned}
\end{equation*}

Choosing $\eta = O(\frac{1}{\sqrt{T}})$ ensures the average squared gradient norm decays as:
\begin{equation*} 
\frac{1}{T} \sum_{t=0}^{T-1} |\nabla F(P_t)|^2 = O\left(\frac{1}{\sqrt{T}}\right). 
\end{equation*}
This completes the convergence proof.

\section{Additional Ablation Study} \label{appendix:additional_ablation}

\bsub{Client Selection Strategy. }
In realistic FL scenarios, clients have heterogeneous hardware and network conditions, making uniform sampling relatively less representative. To this end, we evaluate EcoLoRA under asynchronous FL, where clients update at different speeds. Following FedBuff~\cite{nguyen2022federated}, the server aggregates updates from a buffer of the fastest clients, allowing them to participate more frequently. Results with LLaMA2-7B on Dolly (Table~\ref{tab:client_selection}) confirm that EcoLoRA maintains strong performance even under asynchronous client participation.

\begin{table}[h]
\small
\centering
\begin{threeparttable}
\caption{Comparison of accuracy and parameters (in millions) under different client selection strategies.}
\label{tab:client_selection}
\renewcommand{\arraystretch}{1.1}
\setlength{\tabcolsep}{10pt}
\begin{tabular}{c|c|c|c}
\hline\hline
\textbf{Method} & \textbf{ARC} & \textbf{Upload P.} & \textbf{Total P.} \\
\hline\hline
FedIT & 66.3 & 3523.2 & 6962.5 \\
w/ EcoLoRA & 66.4 & 508.6 & 3955.8 \\
\hline\hline
\end{tabular}
\end{threeparttable}
\end{table}

\bsub{Experiments under Non-IID Conditions with Task Heterogeneity. } In some extreme federated learning scenarios, each client may possess a significantly different data distribution, such as having a distinct task domain. It is important to assess the performance of \sysname under such heterogeneous conditions. We evaluate our method on the Databricks-Dolly-15k dataset by assigning each client a unique task type based on the dataset's category field, using LLaMA-7B as the base model. The results are shown in Table~\ref{tab:comm_cost_comparison}. As shown, \sysname achieves substantial reductions in communication overhead while maintaining competitive performance across non-IID, task-diverse clients.

\begin{table}[h]
\small
\centering
\resizebox{0.48\textwidth}{!}{%
\begin{threeparttable}
\caption{Comparison of accuracy and parameters (in millions) under non-IID conditions divided by task domain. }
\label{tab:comm_cost_comparison}
\renewcommand{\arraystretch}{1.1}
\setlength{\tabcolsep}{2pt}
\begin{tabular}{c|c|c|c}
\hline\hline
\textbf{Method} & \textbf{ARC} & \textbf{Upload Param.} & \textbf{Total Param.} \\
\hline\hline
FedIT                     & 0.664 & 2348.8 & 4697.6 \\
FedIT w/ EcoLoRA          & 0.664 & 285.5  & 2157.3 \\
FLoRA                     & 0.663 & 2181.0 & 23991.4 \\
FLoRA w/ EcoLoRA          & 0.663 & 292.5  & 19105.3 \\
FFA-LoRA                  & 0.665 & 1090.5 & 2181.0 \\
FFA-LoRA w/ EcoLoRA       & 0.666 & 136.8  & 995.0 \\
\hline\hline
\end{tabular}
\end{threeparttable}
}
\end{table}




\end{document}